\documentclass{article}
\usepackage{epsfig}
\input{epsf}
\begin{document}

\title{Stability properties of some perfect fluid
cosmological models}
\author
{Vittorio Gorini$^{1,2}$, Alexander Kamenshchik$^{1,3,4}$, Ugo
Moschella$^{1,2}$,
\\Vincent Pasquier$^{5}$, Alexei Starobinsky$^{4}$}
\date{}
\maketitle \hspace{-6mm}
$^{1}${\it Dipartimento di Scienze
Fisiche e Mathematiche, Universit\`a
dell'Insubria, Via Valleggio 11, 22100 Como, Italy}\\
$^{2}${\it INFN,sez. di Milano, Via Celoria 16, 20133 Milano, Italy}\\
$^{3}${\it Dipartimento di Fisica, Universit\`a di Bologna and
INFN, via Irnerio 46,
40126 Bologna, Italy}\\
$^{4}${\it L.D. Landau Institute for Theoretical Physics,
Russian Academy of Sciences, Kosygin str. 2, 119334 Moscow, Russia}\\
$^{5}${\it Service de Physique Th\'eorique, C.E. Saclay, 91191
Gif-sur-Yvette, France}

\begin{abstract}
Flat FRW perfect fluid cosmologies can be reproduced as particular
solutions of suitable field theoretical models. Here we
investigate the stability of perfect fluid model trajectories with
respect to sets of trajectories of the corresponding field models
having generic initial conditions. It is shown that the
trajectories of barotropic perfect fluid models and those of the
Chaplygin gas model are stable. The total probability to reach the
Chaplygin gas regime early enough to achieve a matter dominated
stage having a realistic duration is calculated for a  scalar
field model assuming a natural measure in the space of initial
conditions taken near a cosmological singularity. An example is
presented of a two-fluid cosmological model where the stability is
partially absent.
\end{abstract}

\section{Introduction}
The discovery of the phenomenon of cosmic acceleration
\cite{accel,accel1} has stimulated an intense activity directed to
the construction of new cosmological models based on different
microscopical models of dark energy producing this acceleration,
in particular, those unifying dark matter and dark energy (see,
e.g., the reviews \cite{Sahni,Padmrep,Ratra}).

It is well-known that for isotropic flat Friedmann-Robertson-Walker
(FRW) cosmological models, for a given
dependence of the cosmological scale factor $a(t)$ on time $t$, it
is always possible to construct a potential $U(\phi)$ for a
minimally coupled scalar field $\phi$ which reproduces this
cosmological evolution exactly (see, e.g., \cite{Barrow,Star}). A
similar statement holds also for tachyon matter (for a detailed
analysis and examples see \cite{we-tach}). If one has an exact
solution arising in a FRW universe filled with a perfect fluid, one
can always implicitly (and often explicitly as well)  find a
scalar field or tachyon model which reproduces this evolution
provided suitable initial conditions are imposed
\cite{Barrow,we-tach}. However, it is necessary to emphasize that
the correspondence between perfect fluid, scalar field and tachyon
cosmological models has a very limited character and amounts
precisely to the existence of identical solutions obtained thanks
to a special choice of initial conditions. If one moves away from
these conditions, the dynamics of a scalar field FRW model can be
very different from that of a corresponding perfect fluid model or
of a model with tachyon matter.

In spite of the existence of a vast variety of scalar field,
tachyon \cite{Sen,tach,Star-tach,Padm,Fein,AF,we-tach},
$k$-essence \cite{k-ess} and other field-theoretical cosmological
models, models based on perfect fluids still retain their own
attractive features due to their simplicity and a smaller number
of parameters.

Consider, for example, the Chaplygin gas cosmological model
introduced in \cite{we-Chap} and further developed in
\cite{Bilic,Fabris,Bento,we-Chap1}. This model is the simplest one
unifying dark matter and dark energy in a FRW background. It is
based on the equation of state

\begin{equation}
p = - \frac{A}{\rho}~, \label{Chap}
\end{equation}
where $A$ is a positive constant. In what follows, we shall
consider only the Chaplygin gas cosmological model with $\rho \geq
\sqrt{A}$ which is equivalent to the energy dominance requirement
that $|p| \leq \rho$. The Chaplygin gas with $\rho < \sqrt{A}$ was
considered elsewhere \cite{phan-Chap,phan-Chap1}. (While this
latter case is obviously not apt to the unification of dark energy
and dark matter, it could be useful in the context of phantom
cosmology \cite{phan-Chap1}).

It was shown in \cite{we-Chap} that the Chaplygin gas cosmological
dynamics can be reproduced using the scalar field model with the
standard kinetic term $\frac{1}{2}\phi_{,\mu}\phi^{,\mu}$ and the
potential
\begin{equation}
U(\phi) = \frac{1}{2}\sqrt{A}\left(\cosh 3\phi + \frac{1}{\cosh
3\phi}\right) \label{pot-Chap}
\end{equation}
(throughout the paper, we use the Landau-Lifshitz sign conventions,
in particular, the space-time signature $(+---)$, and units in which
$c=\hbar=8\pi G/3=1$, then $A = \rho_{min}^2 \ll 1$).
This representation is unique up to a
constant shift of $\phi$. However, the opposite is {\em not} true:
not any (and in fact only one) FRW solution for a minimally coupled
scalar field  with the standard kinetic term and the potential
(\ref{pot-Chap}) satisfies the relation (\ref{Chap}). Thus, there
arises the problem of stability of dynamical trajectories of the FRW
Chaplygin gas model with respect to the family of trajectories of
the scalar field FRW model with the potential (\ref{pot-Chap}).

It has also been noticed  that the cosmological model based on Sen's
effective Lagrangian of tachyon matter \cite{Sen}
\begin{equation}
L = -V(T)\sqrt{1-T_{,\mu}T^{,\mu}} \label{Sen}
\end{equation}
with the potential $V(T) = \sqrt{A}$ exactly coincides with the
Chaplygin gas model \cite{Star-tach}. This shows that the
representation (\ref{pot-Chap}) is not the only possible
microscopical model for the Chaplygin gas if the assumption of the
standard (minimal) kinetic term is dropped. Moreover, remarkably,
the tachyon representation of the Chaplygin gas has {\em no} extra
solutions with a time-like gradient: all solutions of the model
(\ref{Sen}) with $T_{,\mu}T^{,\mu} > 0$ satisfy the equation of
state (\ref{Chap}).

It is known that the Chaplygin gas model has serious problems with
linear perturbations on the FRW background: analysis shows that it
has a wrong transfer function for matter perturbations \cite{STZW}
and a too low value of high multipole temperature anisotropy
\cite{CF,AFBC,BD}. Account of non-linear effects, while leading to
some new features, is not sufficient to save the simple Chaplygin
gas model \cite{BLTV}.
Nevertheless, we believe that in spite of this failure, meaning
only that this model is too simplified to describe the real
Universe, the very idea of dark matter and dark energy unification
remains alive.
Models with a more complicated effective equation of
state resulting in a lower value for the sound velocity at recent
times may avoid this difficulty.
In particular, such models may be based on Lagrangians of the
type (\ref{Sen}) but with different dependence on $T_{,\mu}T^{,\mu}$
\cite{Scherrer}.
Also, non-adiabatic perturbations in the Chaplygin gas \cite{Reis}
and more general models \cite{Reis1} open a new possibility to
cure this problem.
Thus, it seems to us  reasonable
to share the hope, expressed by Bilic et al \cite{BLTV},
that ``the Chaplygin gas model is not so much wrong as
incomplete''. For this reason, in this paper we restrict ourselves
to the FRW background in order to investigate if the Chaplygin gas
model has problems already at this level, as was recently claimed
in \cite{Perrotta}.

It is tempting to conceive that the appearance of the Chaplygin
gas in cosmology has a microscopic origin and represents some
remnant of a more fundamental theory (see, for example,
\cite{Bilic1,we-review}), while it is difficult to imagine that
the scalar field model with  potential (\ref{pot-Chap}) has a
special meaning. However, just this scalar field representation of
 the Chaplygin gas model was used in \cite{Perrotta}. Here a
numerical study was performed of the problem of stability of the
Chaplygin gas model trajectories in the framework of a wider set
of trajectories of the scalar field model (\ref{pot-Chap}). This
numerical analysis was used to draw the conclusion that these
trajectories are unstable, which was interpreted as a new and very
serious problem for the Chaplygin gas cosmology arising already at
the level of background solutions. As we have already noted above,
the scalar field model (\ref{pot-Chap}) is neither unique, nor a
fundamental representation of matter with the Chaplygin gas model
equation of state. So, the problem of stability of the Chaplygin
gas model trajectories with respect to the set of trajectories of
this model cannot serve as a touchstone for the validity of the
Chaplygin gas model as a candidate for a model unifying dark
energy and dark matter. In particular, there is no such problem at
all for the tachyon representation (\ref{Sen}).

However, the problem of stability of trajectories of perfect fluid
models with respect to a set of trajectories of underlying
field-theoretical models is of interest by its own. That is why we
study this problem analytically in this paper. It will be shown
that the Chaplygin gas model cosmological evolution is
{\it stable} with respect to  small deviations, considered in the
linear approximation. Then we consider other perfect fluid models
and their scalar field and tachyon ``relatives'' and show that the
linear stability of perfect fluid trajectories is a non-trivial
phenomenon which does not always occur.

The method to study stability of perfect fluid trajectories is the
following one. We introduce a parameter measuring the deviation of a
scalar field or tachyon model trajectory from the corresponding
perfect fluid trajectory and then study the time evolution of this
parameter in the neighbourhood of the perfect fluid trajectory in
the linear approximation (namely, to first order in the parameter).
In Sec. II this method will be applied to
different perfect fluid models
in the regime in which deviations between scalar field and perfect fluid
trajectories are small and, therefore, can indeed be treated to first
order.
Section III will be devoted to the
detailed study of the Chaplygin gas model. Here we shall abandon
the hypothesis of
small linear deviations. The total
probability to reach the Chaplygin gas regime early enough to
achieve a matter dominated stage having a realistic duration will be
calculated for a  scalar field model assuming a natural measure in
the space of initial conditions taken near a cosmological
singularity. It will be shown that the Chaplygin gas trajectory is
rather stable not only in a local but also in a global sense.

\section{Linear analysis of the stability of perfect fluid cosmologies}
The first model that we investigate is the Chaplygin gas. We
introduce the parameter
\begin{equation}
x \equiv \rho p + A, \label{param}
\end{equation}
which is equal to zero for the Chaplygin gas cosmological trajectory. Now,
consider the scalar field model with the potential (\ref{pot-Chap}). For
this model, the Einstein equations for a FRW space-time with the zero
spatial curvature have the form
\begin{equation}
h^2\equiv \left({\dot a(t)\over a(t)} \right)^2 =
\rho = \frac{\dot{\phi}^2}{2} + U(\phi), \label{energy}
\end{equation}
\begin{equation}
\dot h = -{3\over 2} (\rho +p)~, ~~~p = \frac{\dot{\phi}^2}{2} - U(\phi).
\label{pressure}
\end{equation}
Substituting the expressions (\ref{energy}) and (\ref{pressure})
into the formula (\ref{param}), one gets
\begin{equation}
x = \frac{\dot{\phi}^4}{4} - U^2(\phi) + A. \label{param1}
\end{equation}
The time derivative of $x$ is
\begin{equation}
\dot{x} = \ddot{\phi} \dot{\phi}^3 - 2UU'\dot{\phi}, \label{time}
\end{equation}
where $U' \equiv dU/d\phi$. Using the Klein-Gordon and Friedmann
equations, one has
\begin{equation}
\ddot{\phi} = -3\sqrt{\frac{\dot{\phi}^2}{2}+U} \dot{\phi} - U'.
\label{KGF}
\end{equation}
Substituting (\ref{KGF}) into Eq. (\ref{time}), we obtain
\begin{equation}
\dot{x} = -3\dot{\phi}^4\sqrt{\frac{\dot{\phi}^2}{2}+U}
-\dot{\phi}^3 U' - 2UU'\dot{\phi}. \label{time1}
\end{equation}
Now, using Eq. (\ref{param1}), we can express the time derivative
of the scalar field as
\begin{eqnarray}
&&\dot{\phi} = -\sqrt{2}(U^2-A+x)^{1/4}\nonumber \\
&&=-\sqrt{2}\left(\frac{A}{4}\tanh^2 3\phi + x\right)^{1/4}.
\label{phi-dot}
\end{eqnarray}
Substituting this expression together with the explicit form of
the potential (\ref{pot-Chap}) into Eq. (\ref{time1}), we get the
expression for the time derivative $\dot{x}$ as a function of the
field $\phi$ and the parameter $x$. Precisely:
\begin{eqnarray}
&&\dot{x} = -3(A\tanh^2 3\phi + 4x)\nonumber \\
&&\times\left[\frac{1}{2}\sqrt{A}\left(\cosh 3\phi +
\frac{1}{\cosh 3\phi}\right) + \left(\frac{A}{4}\tanh^2 3\phi +
x\right)^{1/2}\right]^{1/2}\nonumber \\
&&-3\sqrt{2A}\frac{\sinh^3 3\phi}{\cosh^2
3\phi}\left(\frac{A}{4}\tanh^2 3\phi + x\right)^{3/4}\nonumber \\
&&+\frac{3\sqrt{2}A}{2}\frac{\sinh^3 3\phi}{\cosh^2 3\phi}
\left(\cosh 3\phi + \frac{1}{\cosh 3\phi}\right)
\left(\frac{A}{4}\tanh^2 3\phi + x\right)^{1/4}.
\label{time-exact}
\end{eqnarray}
We write down the leading term of the  expansion of the right-hand
side of Eq. (\ref{time-exact}) in the neighborhood of the point $x
= 0$, disregarding terms of order higher than one:
\begin{equation}
\dot{x} = -\frac{3A^{1/4}(5\sinh^2 3\phi + 6)}{2(\cosh
3\phi)^{3/2}} x + o(x). \label{time20}
\end{equation}
The sign of the coefficient of the linear term in $x$ is negative.
Therefore, trajectories of the Chaplygin gas cosmological model,
if considered as a subset of trajectories of the scalar field
model (\ref{pot-Chap}), are stable.

Next we calculate the rate of convergence of scalar field model
trajectories to the Chaplygin gas solution $x=0$. For the Chaplygin gas
the value of the Hubble parameter is
\begin{equation}
h = A^{1/4}\sqrt{\cosh 3\phi}. \label{hubble-chap}
\end{equation}
Thus, the relation between the coefficient $\gamma$ from the
equation
\begin{equation}
\dot{x} = -\gamma x + o(x) \label{gamma-def}
\end{equation}
and the Hubble parameter is the following:
\begin{equation}
\frac{\gamma}{h} = \frac{3}{2}\left(5 + \frac{1}{\cosh^2
3\phi}\right) \label{rel-Chap}
\end{equation}
and hence
\begin{equation}
7.5 \leq \frac{\gamma}{h} \leq 9. \label{rel-Chap1}
\end{equation}
Thus, the Chaplygin gas behaviour is quickly achieved once $|x|
\stackrel{<}{\sim} A$.

 For the tachyon matter representation (\ref{Sen}) of the Chaplygin
gas model ($V(T)=\sqrt A =const$), the variable $x$ is identically
zero, so there is no problem of stability at all, as has been
pointed out above.

We now consider  another well known perfect fluid model obeying
the barotropic equation of state
\begin{equation}
p = k\rho~. \label{barotrop}
\end{equation}
A spatially flat FRW universe filled with this fluid expands
according to the power law
\begin{equation}
a(t) = a_0 t^{\frac{2}{3(1+k)}}. \label{power}
\end{equation}
The cosmological dynamics (\ref{power}) can be reproduced in the
scalar field model with the standard kinetic term and the
potential \cite{scalar,we-tach}
\begin{equation}
U(\phi) = U_0\exp(3\sqrt{1+k}\phi). \label{pot}
\end{equation}
In order to measure the deviation of a trajectory of the model
(\ref{barotrop}) from that of the model (\ref{pot}), we introduce
a variable $x$ as
\begin{equation}
x \equiv \frac{p}{\rho} - k. \label{param-bar}
\end{equation}
Following the method described above for the Chaplygin gas model,
we come to the equation
\begin{eqnarray}
&&\dot{x} = {3\over
2}\sqrt{2U(1-(k+x)2)}(\sqrt{1+k}-\sqrt{1+k+x})
\nonumber \\
&&=-{3\over 2}\sqrt{2U(1-k)}x + o(x) \label{time3}
\end{eqnarray}
and see that the perfect fluid trajectory is stable in this case,
too.

 For the power-law evolution (\ref{power})
\begin{equation}
h = \sqrt{\frac{2U}{1-k}} \label{hubble-power}
\end{equation}
and
\begin{equation}
\frac{\gamma}{h} = \frac{3}{2}(1-k). \label{rel-power}
\end{equation}
Thus, we have a convergence of trajectories of the scalar field
model to that of the perfect fluid model, whose rapidity depends
on the value of $k$.

It is known that in the case of negative pressure $k < 0$, the
cosmological evolution given by Eq. (\ref{power}) can be
reproduced in the framework of the tachyon model (\ref{Sen})
\cite{Padm,Fein,AF} with the potential
\begin{equation}
V(T) = \frac{4\sqrt{-k}}{9(1+k)T^2}. \label{PF}
\end{equation}
Using the known expression for the tachyon energy density and
pressure
\begin{equation}
\rho = \frac{V(T)}{\sqrt{1-\dot{T}^2}}, \label{en-tach}
\end{equation}
\begin{equation}
p = - V(T)\sqrt{1-\dot{T}^2}, \label{pres-tach}
\end{equation}
one can easily see that
\begin{equation}
x = -(1-\dot{T}^2) - k. \label{x-tach}
\end{equation}
Using the tachyon field equation
\begin{equation}
\ddot{T} + 3\sqrt{V}(1-\dot{T}^2)^{3/4}\dot{T}
+(1-\dot{T}^2)\frac{V,_{T}} {V} = 0, \label{tach-KG}
\end{equation}
one can show that
\begin{equation}
\dot{x} = -\frac{1-k}{(1+k)T}x + o(x), \label{tach-time}
\end{equation}
and, hence, the perfect fluid trajectory (\ref{power}) is stable
with respect to the tachyon model having the potential (\ref{PF}).
As far as the case $k > 0$ is concerned, it was shown in
\cite{we-tach} that the dynamics of the perfect fluid model can be
reproduced in the framework of the Born-Infeld action of the type
\begin{equation}
L = W(T) \sqrt{T_{,\mu}T^{,\mu}-1}, \label{BI}
\end{equation}
with the potential
\begin{equation}
W(T) = \frac{4\sqrt{k}}{9(1+k)T^2}. \label{BI-pot}
\end{equation}
A similar analysis shows that the trajectory (\ref{power}) is
stable also with respect to the set of trajectories of the model
(\ref{BI}), (\ref{BI-pot}).

The value of the Hubble parameter for these simple tachyon models is
\begin{equation}
h = \frac{2}{3\sqrt{1+k}T} \label{hubble-tach}
\end{equation}
and, hence,
\begin{equation}
\frac{\gamma}{h} = \frac{3(1-k)}{2\sqrt{1+k}}. \label{rel-tach}
\end{equation}
Thus, the rate of convergence depends on the value of the
parameter $k$. Note that in contrast to  formula
(\ref{rel-power}), the relation $\frac{\gamma}{h}$ in
(\ref{rel-tach}) is not bounded above and tends to infinity when
$k \rightarrow -1$.

We now consider a more complicated family of cosmological models.
In  paper \cite{we-tach}, a two-fluid cosmological model was
studied. One of this fluids is merely a cosmological constant $p =
-\rho = -\Lambda$, while the second fluid obeys the equation of
state $p = k\rho$, where $-1 < k < 1$. The cosmological evolution
of the model is described by the formula
\begin{equation}
a(t) = a_0 \left(\sinh
\frac{3\sqrt{\Lambda}(1+k)t}{2}\right)^{2/[3(1+k)]}.
\label{our-evol}
\end{equation}

It has been shown in \cite{we-tach} that the cosmological dynamics
(\ref{our-evol}) can be reproduced in the scalar field model with
the standard kinetic term and the potential
\begin{equation}
U(\phi) =
\Lambda\left(1+\frac{1-k}{2}\sinh^2\frac{3\sqrt{1+k}\phi}{2}
\right), \label{pot-our}
\end{equation}
or in the tachyon model with the potential
\begin{eqnarray}
&&V(T) =
\frac{\Lambda}{\sin^2\left[\frac32\sqrt{\Lambda(1+k)}T\right]}
\nonumber \\
&&\times\sqrt{1-(1+k)\cos^2\left[\frac32\sqrt{\Lambda(1+k)}T\right]}.
\label{pot-tach}
\end{eqnarray}

The parameter defining a deviation of the cosmological dynamics of
the field-theoretical models (\ref{pot-our}), (\ref{pot-tach})
from the two-fluid behaviour (\ref{our-evol}) looks as follows:
\begin{equation}
x = \frac{p + \Lambda}{\rho - \Lambda} - k. \label{param-our}
\end{equation}
 For the scalar field model with the potential (\ref{pot-our}), the
time derivative of the parameter $x$ in the neighbourhood of the
point $x = 0$ is
\begin{equation}
\dot{x} =
-\frac{3\sqrt{\Lambda}(1-k)}{2\cosh\frac{3\sqrt{1+k}\phi}{2}}
\left[\cosh^2\frac{3\sqrt{1+k}\phi}{2} - \frac{1+k}{1-k}\right]x +
o(x). \label{time-our}
\end{equation}
One can see that in the case $k < 0$, the coefficient of $x$ in
the right-hand side of Eq. (\ref{time-our}) is always negative.
Hence, the trajectory (\ref{our-evol}) is stable. If the parameter
$k$ is positive, the trajectory (\ref{our-evol}) is stable for
\begin{equation}
\phi > \frac{2}{3\sqrt{1+k}}{\rm arccosh} \sqrt{\frac{1+k}{1-k}}
\label{condition}
\end{equation}
and is unstable when the condition (\ref{condition}) is not
satisfied. Thus, the trajectory (\ref{our-evol}) becomes unstable
close to the de Sitter point $\phi = 0, \dot{\phi} = 0, \dot{a}/a
= \sqrt{\Lambda}$. The condition (\ref{condition}) could be also rewritten
as
\begin{equation}
\Omega_{\Lambda} < \frac{1-k}{1+k},
\label{condition11}
\end{equation}
where $\Omega_{\Lambda}$ denotes, as usual,
the ratio of the cosmological constant
to the general energy density in two-fluid model.

As regards  the tachyon model (\ref{pot-tach}), using
expressions (\ref{en-tach}),(\ref{pres-tach}) and (\ref{tach-KG}),
after cumbersome but straightforward calculations we come to the
following equation:
\begin{equation}
\dot{x} = \frac{3\left[2k(1-k)+(k+1)2\cos^2
\left[\frac32\sqrt{\Lambda(1+k)}T\right]
\sin^2\left[\frac32\sqrt{\Lambda(1+k)}T\right]\right]}
{2\sin\left[\frac32\sqrt{\Lambda(1+k)}T\right]\times
\left[(1-k)-(1+k)\cos^2\left[\frac32\sqrt{\Lambda(1+k)}T\right]\right]}x
+ o(x). \label{time-tach-our}
\end{equation}
If $k < 0$, the denominator of the coefficient of $x$ at the right-hand
side of Eq. (\ref{time-tach-our}) is positive while the numerator
is negative at
\begin{equation}
\cos^2\left[\frac32\sqrt{\Lambda(1+k)}T\right] > \frac12 +
\sqrt{\frac14 + \frac{2k(1-k)}{(k+1)^2}} \label{condition1}
\end{equation}
or at
\begin{equation}
\cos^2\left[\frac32\sqrt{\Lambda(1+k)}T\right] < \frac12 -
\sqrt{\frac14 + \frac{2k(1-k)}{(k+1)^2}}. \label{condition2}
\end{equation}
Thus, the perfect fluid model trajectory (\ref{our-evol}) will be
stable if either one of the conditions (\ref{condition1}) or
(\ref{condition2}) is satisfied. If
\begin{equation}
k < \frac{5-4\sqrt{2}}{7}, \label{condition3}
\end{equation}
then trajectory (\ref{our-evol}) is stable for any value of the
tachyon field $T$.

In the case $k > 0$, the numerator of the coefficient of $x$ in the
left-hand side of Eq. (\ref{time-tach-our}) is always positive and
the denominator is negative if
\begin{equation}
\cos^2\left[\frac32\sqrt{\Lambda(1+k)}T\right] > \frac{1-k}{1+k}.
\label{condition4}
\end{equation}
Thus, the perfect fluid trajectory is unstable in the vicinity of
the de Sitter point $\cos\left[\frac32\sqrt{\Lambda(1+k)}T\right]
= 0, \dot{T} = 0, \dot{a}/a = \sqrt{\Lambda}$.

For the tachyon two-fluid model we have
\begin{equation}
h = \frac{\sqrt{\Lambda}}{\sin \left(\frac32 \sqrt{\Lambda(1+k)}T\right)}
\label{Hubble-two-tach}
\end{equation}
and
\begin{equation}
\frac{\gamma}{h} = \frac{3\left((2k(1-k)+(1+k)2 \cos^2 \left(\frac32
\sqrt{\Lambda(1+k)}T \right)\sin^2 \left(\frac32 \sqrt{\Lambda(1+k)}T\right)
\right)}
{2\left(((1+k)\cos^2 \left(\frac32 \sqrt{\Lambda(1+k)}T\right) - (1-k)\right))}
\label{rel-two-tach}
\end{equation}
and different regimes of convergence are possible depending on
values of $k, \Lambda$ and $T$.

\section{Analysis of a large class of initial conditions for the
Chaplygin gas}
In this section we go beyond the linear analysis, developed in the
preceding section, of the small deviations of the trajectories of
scalar field and tachyon cosmological models from the trajectories
of the corresponding perfect fluid models. For the most interesting
case of the Chaplygin gas, we consider a broad variety of initial
conditions for the corresponding scalar field model. However,
in order to
obtain definite results,  we need first to specify which models
are compared to each other. For realistic cosmological models
describing not only the present Universe, but the early Universe
too, one has to include radiation. So, a possible perfect fluid model
is the two component one consisting of the Chaplygin gas and
radiation ($p_{\gamma}=\rho_{\gamma}/3$); we neglect baryons here.
But with which scalar field model should we compare it with?
If we want to
introduce a minimally coupled scalar field $\phi$ with the standard
kinetic term and some potential $U(\phi)$ which models the Chaplygin
gas only (so that the comparison model is  scalar field +
radiation), this potential should be very peculiar: it becomes
strongly divergent at some finite value of $\phi$.

Indeed, though in this case we cannot obtain an exact form of this
potential, we can study its asymptotic behavior for
$\rho,\rho_{\gamma}\gg \sqrt{A}$ and match it to the potential
(\ref{pot-Chap}) valid for $\rho \gg \rho_{\gamma}$. In the former
regime, the Chaplygin gas may be well approximated by dust-like
matter. So, let us consider a scalar field which mimics only dust
in a universe filled with dust and radiation. As usual, the scalar
field
potential is
\begin{equation}
U = \frac12(\rho - p) = \frac{C}{2a^3},
\label{potential}
\end{equation}
where $C$ is a constant describing the amount of dust.
Then
\begin{equation}
\dot{\phi}^2 = \rho + p = \frac{C}{a^3},
\label{kinetic}
\end{equation}
or,
\begin{equation}
\dot{\phi}^2 = \left(\frac{d\phi}{da}\right)^2 \frac{\dot{a}^2}{a^2} a^2,
\label{kinetic1}
\end{equation}
where the squared Hubble parameter is defined from the Friedmann
equation which contains both dust and radiation:
\begin{equation}
\frac{\dot{a}^2}{a^2} = \frac{C}{a^3} + \frac{D}{a^4}.
\label{Hubble-Fr}
\end{equation}
Substituting (\ref{Hubble-Fr}) into Eqs. (\ref{kinetic}), (\ref{kinetic1}),
one obtains
\begin{equation}
\frac{d\phi}{da} = \frac{1}{\sqrt{a(a+a_{eq})}},
\label{int}
\end{equation}
where
\begin{equation}
a_{eq} \equiv \frac{D}{C}
\label{a0-def}
\end{equation}
is the value of the scale factor when dust and radiation have equal
energy densities. Introducing a new variable
\begin{equation}
a = a_{eq} \sinh^2 \chi,
\label{chi-def}
\end{equation}
one can integrate Eq. (\ref{int}) obtaining
\begin{equation}
\phi = 2\chi+\phi_0 = 2 {\rm arcsinh} \sqrt{\frac{a}{a_{eq}}} +\phi_0.
\label{int1}
\end{equation}
Inverting the relation (\ref{int1}), one has
\begin{equation}
a = a_{eq} \sinh^2 \frac{\phi-\phi_0}{2}.
\label{inverse}
\end{equation}
Substituting into Eq. (\ref{potential}), it gives
the potential which as announced before strongly diverges at
$\phi = \phi_0$
\begin{equation}
U(\phi) = \frac{C}{2a_{eq}^3 \sinh^6\frac{\phi-\phi_0}{2}}.
\label{poten1}
\end{equation}

On the other hand, if the Chaplygin gas dominates over radiation
($\rho \gg \rho_{\gamma}$), the potential $U(\phi)$ should have
the form (\ref{pot-Chap}). Let us match the expressions
(\ref{pot-Chap}) and (\ref{poten1}) at the matter dominated stage
$\rho\gg \sqrt A,~\rho \gg \rho_{\gamma}$. The energy density of
the Chaplygin gas has the following dependence on $a(t)$: $\rho =
\sqrt{A+C^2/a^6}$ where the constant $C$ is the same as in Eq.
(\ref{Hubble-Fr}). Let $\Omega_{\gamma}\ll 1$ be the radiation
energy density at the ``present'' moment $a=a_0,\, \rho = \rho_0 =
\sqrt{A/0.7},\, \dot a/a =H_0=\sqrt{\rho_0}$, i.e., at the moment
when $|p_{tot}|/\rho_{tot}$ has the same value $0.7$ as in the
standard $\Lambda$CDM model with $\Omega_m=0.3,\,
\Omega_{\Lambda}=0.7$. Then
\begin{eqnarray}
{C^2\over a_0^6}=\rho_0^2 - A\approx 0.43 A~,~~D=\Omega_{\gamma}
\rho_0a_0^4~, \\
{a_{eq}\over a_0}\equiv {1\over 1+z_{eq}}=
\Omega_{\gamma}\, {\rho_0\over \sqrt{\rho_0^2 - A}}\approx
1.83\Omega_{\gamma}~.
\end{eqnarray}
In the region $1\ll \phi \ll \phi_0$,
\begin{equation}
U(\phi)= {1\over 4}\sqrt A \, e^{3\phi}= {32C\over a_{eq}^3}\,
e^{3\phi-3\phi_0}~,
\end{equation}
where the first expression for $U(\phi)$ is the asymptote of
(\ref{pot-Chap}) for $\phi \gg 1$, while the second one is the
asymptote of (\ref{poten1}) for $\phi_0 - \phi \gg 1$. Therefore,
\begin{equation}
e^{3\phi_0}={128 C\over a_{eq}^3\sqrt{A}}\approx {13.8\over
\Omega_{\gamma}^3}~, ~~ \phi_0\approx |\ln \Omega_{\gamma}|
+0.87\gg 1~.
\end{equation}
Thus, inclusion of radiation drastically changes the scalar field
potential from having an exponential growth at infinity to having
 a pole at finite value of $\phi$.

If, on the other hand, we choose to model {\em both} the
Chaplygin gas and radiation by one scalar field, then $U(\phi)$
is regular for  finite $\phi$, but the model itself becomes
artificial since, in particular, it misses the fact that radiation
has a non-zero entropy. Also, in that case we would investigate
the stability of a fluid model different from the original
Chaplygin gas one. Note once more that no such ambiguity arises
in the case of the tachyon representation (\ref{Sen}) of the
Chaplygin gas.

So, to present methods able to deal with the case of large
deviations between trajectories of fluid and field models, and to
get quantitative results about the probability measure of favourable
trajectories, we restrict ourselves to the case of the pure
Chaplygin gas model compared to the scalar field model with the
potential (\ref{pot-Chap}), both applied up to arbitrary large
curvatures. It is less complete from the physical point of view, but
more well defined instead.

Then, the second step is to decide which solutions of the scalar
field model we consider as favourable ones. A natural choice for this
toy model is to choose all those solutions for which the Chaplygin
gas attractor has been already reached by the moment of the
matter-radiation equality in the real Universe $z=z_{eq} \approx 3450$
(the value from the WMAP data \cite{WMAP} is used here). In
particular, it may be reached earlier. Note that the redshift $z
\sim 1$ should correspond to the moment when $\rho \sim \sqrt A$.

We consider the case $x \gg A$ which is typical at
sufficiently early time near the singularity. In this regime, the
kinetic energy of the scalar field much exceeds its potential energy.
Then
\begin{equation}
a(t)\propto t^{1/3}~, ~~~\phi =\pm {\sqrt 2\over 3}\ln{1\over t} +
const~. \label{free}
\end{equation}
Thus, for $|\phi|\gg 1$, the potential $U(\phi)$ decreases $\propto
\exp(3|\phi|)\propto t^{-\sqrt 2}$ in the course of expansion, if $\phi$ and
$\dot{\phi}$ have opposite signs (in the opposite case $U(\phi)$ even
increases $\propto t^{\sqrt{2}}$),
i.e., more slowly than the kinetic energy which is $\propto t^{-2}$.
Therefore, $U$ generically becomes of the order of the kinetic
energy at some characteristic moment of time $t=t_U$.

If $|\phi_U|\equiv |\phi(t_U)|\gg 1$ at this moment, the long
dust-like 'tracking' regime follows with
\begin{equation}
a(t)\propto t^{2/3}~, ~~~\phi = \pm {2\over 3}\ln{1\over t} +
const = \mp\ln a + const, ~~~ U={\dot\phi^2\over 2}~. \label{dust}
\end{equation}
The latter regime just represents the behaviour of the Chaplygin
gas model for large $\rho$. Thus, for generic initial conditions
defined at some sufficiently large redshift $z\gg z_{eq} \approx
3450$ (the value from the WMAP data \cite{WMAP} is used here for
the redshift $z_{eq}$ of the matter-radiation equality), the only
necessary and sufficient condition for the scalar field model to
reach the Chaplygin gas trajectory at a redshift exceeding
$z_{eq}$ is
\begin{equation}
|\phi_U| > \ln z_{eq} \approx 8 \label {cond}
\end{equation}
(recall that $\phi$ is measured in units of $(3/8\pi G)^{1/2}$).
The quantity $\phi_0$ where the potential $U$ diverges in
Eq. (\ref{poten1}) is of order of $\ln z_{eq}$, too.
Also, notice that in the case $x \ll -A$, the potential energy is
much higher than the kinetic energy.
A simple analysis of the Klein-Gordon equation shows that in this
case the scalar field $\phi$ is in the fast rolling regime and,
after a Hubble time interval $\sim h^{-1}$, it has no alternative
but to follow the tracking attractor (\ref{dust}) arriving again
at the Chaplygin-like regime. It is clear that, for natural initial
conditions defined near a cosmological singularity when $\rho$ is
of the order of unity (i.e., of the Planck value in usual units),
this condition is satisfied for an open set of initial conditions
with non-zero measure, being therefore generic. It is instructive
to compare the value (\ref{cond}) to a similar quantity which
arises in the theory of chaotic inflation: an initial value of a
homogeneous inflaton scalar field with a polynomial potential at
the beginning of inflation which is required to have more than 60
e-folds of inflation. For the simplest case $U=m^2\phi^2/2$, the
standard expression for the number of e-folds leads to
$|\phi_{in}| > ({4\over 3}\cdot 60)^{1/2}\approx 9$. Thus,
condition (\ref{cond}) is even less restrictive than the one which
is assumed in the chaotic inflationary scenario. On the other
hand, two of the three specific initial conditions (taken at the
redshift $z\sim 10^5$) studied in  \cite{Perrotta} do not satisfy
(\ref{cond}). The third one -- their case (a) -- does  satisfy
(\ref{cond}) and embodies a model where the Chaplygin gas
behaviour is valid from sufficiently early time. In \cite{Perrotta},
it was considered as a bad one for the Chaplygin gas model
stability because it does not satisfy the final condition
$\rho \sim \sqrt{A}$ at $z=0$.
However, for given initial values of $\phi$ and
$\dot\phi$, the redshift $z$ may not be taken arbitrarily; it
should be counted back in time from the present moment when $z=0$,
as has been pointed out above.
With this re-interpretation and proper normalization of $z$, this
case becomes favorable for the Chaplygin gas model stability.

For illustrative purposes, we have performed  some numerical
calculations integrating Eqs. (\ref{energy}) and (\ref{KGF}). We 
have studied the most interesting case of the transition 
to the Chaplygin gas behavior in the regime 
f$\rho\gg \sqrt A$. 
In these conditions, the Chaplygin gas behavior is practically 
indistinguishable from the dust like one.
We plot the time dependence of the
quantity $x/h^4$ which is equal to $w + A/h^4$ where $w\equiv p/\rho$
is the equation of state parameter. Thus, it just
reduces to $w$ in the most interesting regime $\rho\gg \sqrt A$. 
This quantity 
is equal to $1$ in the kinetic dominated regime and to $-1$ in the
potential dominated regime with $\rho\gg \sqrt A$. On the other
hand, the condition $|x/\rho^2|\ll 1$ is just the condition that
we are close to the Chaplygin gas behavior for any $\rho$.

The "realistic" value of $A$ is given by the condition $\rho_{min}=
\sqrt A = 0.7H_0^2$ where $H_0$ is the Hubble constant. The value 
$H_0=70$ km s$^{-1}$ Mpc$^{-1}$ corresponds to $H_0=3.5\cdot 10^{-61}$ 
in our units. Therefore, $A\approx 10^{-242}$, while natural initial 
conditions have to be placed somewhere near $t\sim 1$. In order to avoid 
long numerical runs, we propose the following
scheme. Let us rescale time to $\tilde t = 10^{53}t$, the energy
density to $\tilde \rho = 10^{-106}\rho$ and $\tilde x =
10^{-212}x$. This means that time is measured in units of $\sim
1.56\times 10^{10}$ s, i.e., roughly 100 times less than the
matter-radiation energy equality moment. Then the rescaled value of A is
$10^{212} \times 10^{-242}=10^{-30}$. All tildas are omitted below.

Now let us take different initial conditions for $\phi$ and $\dot\phi$
on the circle $h=\dot a/a = 1$ at the initial moment chosen 
conventionally at $t_0=1$ , (strictly speaking, this time moment
should be shifted by some amount less than 1 depending on initial
conditions to get an initial singularity strictly at $t=0$, but it
is not important for $t\gg 1$).

The initial conditions which we consider are the following ones:\\
a)$\dot{\phi}_0=0, U_0=1, \phi_0= 11.98$;\\
b)$\dot{\phi}_0=1, U_0=0.5, \phi_0=11.51$;\\
c)$\dot{\phi}_0=-1, U_0=0.5, \phi_0=11.51$;\\
d)$\dot{\phi}_0=1.342, U_0=0.1, \phi_0=11.21$;\\
e)$\dot{\phi}_0=-1.342, U_0=0.1, \phi_0=11.21$;\\
f)$\dot{\phi}_0=1.378, U_0=0.05, \phi_0=10.98$;\\
g)$\dot{\phi}_0=-1.378, U_0=0.05, \phi_0=10.98$.

 For all the above cases we have made a run up to $t=1000$ to show that the
Chaplygin gas behavior ($|x|/\rho^2$ becomes much less than $1$)
has been reached by $t\sim 100$, i.e., just when the matter
dominated stage begins in the real Universe. The dependence of
$x/h^4$ is plotted in Fig. 1. For all the cases we have calculated
also the value of the scalar field $\phi_U$ at the moment $t_U$
which is approximately equal to $100$ for the cases a)-f). The corresponding
values of $\phi_U$ are the following ones:\\
a)$\phi_U = 8.416$;\\
b)$\phi_U = 8.422$;\\
c)$\phi_U = 8.389$;\\
d)$\phi_U = 8.425$;\\
e)$\phi_U = 8.287$;\\
f)$\phi_U = 8.429$.\\
Apparently, all these values satisfy the criterion (\ref{cond}). The case 
g) represents the borderline where the condition (\ref{cond}) is saturated. 
In this case the moment $t_U \approx 130$ and $\phi_U = 8.015$. Thus, it 
is not possible to take initial values of $U_0$ smaller than 
$0.05$ for $\dot \phi_0<0$. 

Therefore, we see that with these "late" initial conditions, it is enough 
to have a sufficiently large $\phi_0$ or $U(\phi_0)$ to get a long dust-like 
stage described by the Chaplygin gas. It will be shown below that for 
"natural" initial conditions defined at the Planck scale, this requirement 
may be further relaxed. In other words, the earlier we take an initial moment, 
the larger is the "good" region in the space of initial conditions. 

%\newpage
\begin{figure}[h]
\epsfxsize6.5cm \epsfbox{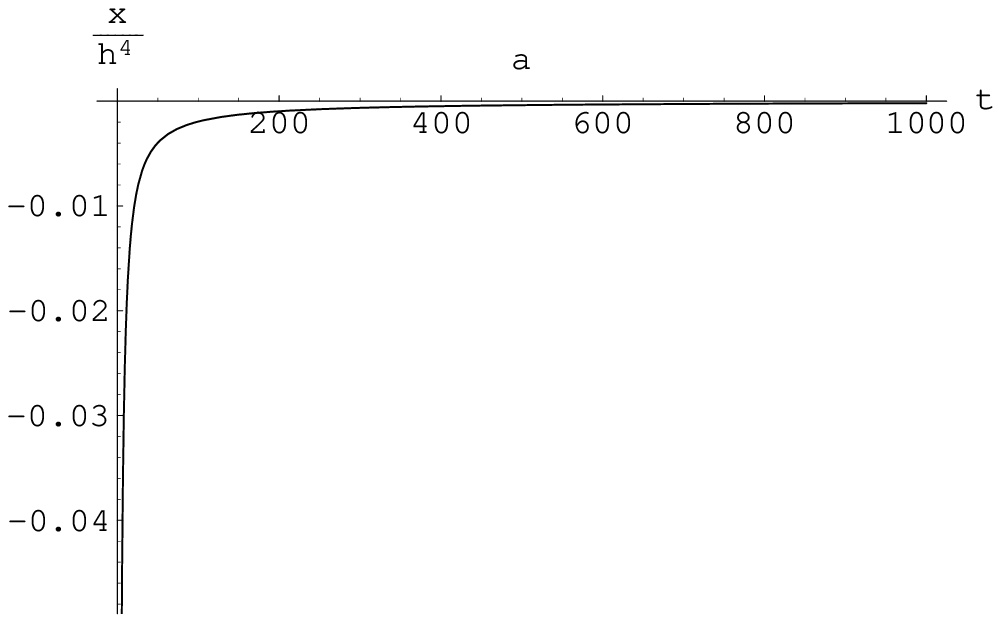} \epsfxsize6.5cm
\epsfbox{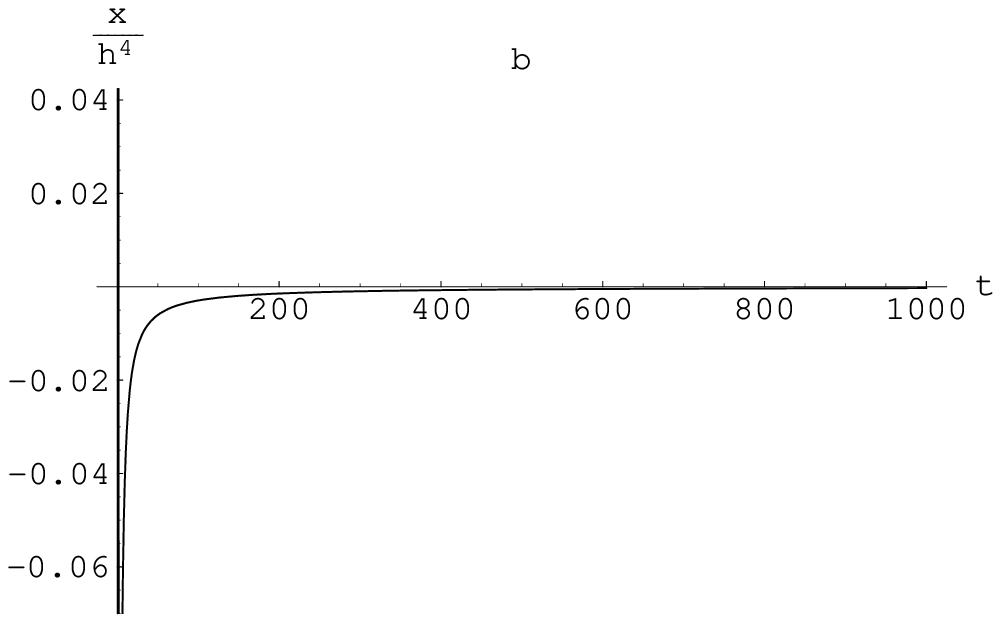} \epsfxsize6.5cm \epsfbox{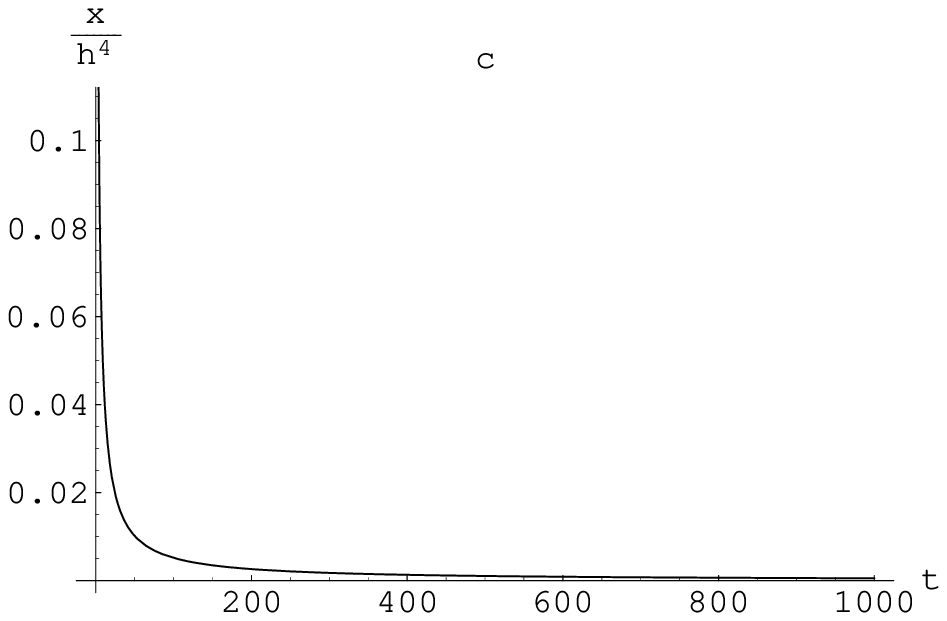}
\epsfxsize6.5cm \epsfbox{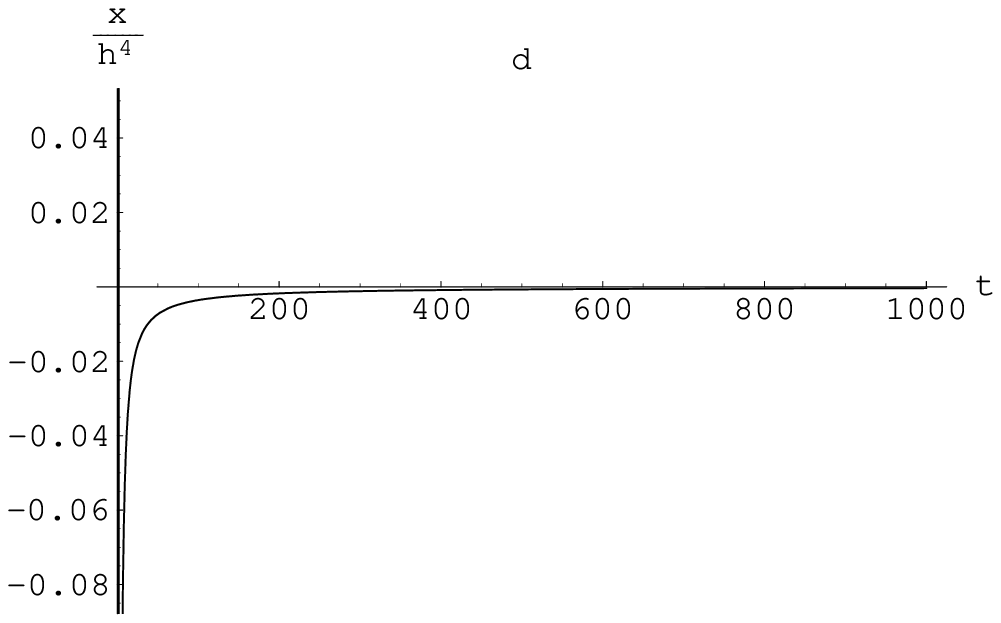} \epsfxsize6.5cm
\epsfbox{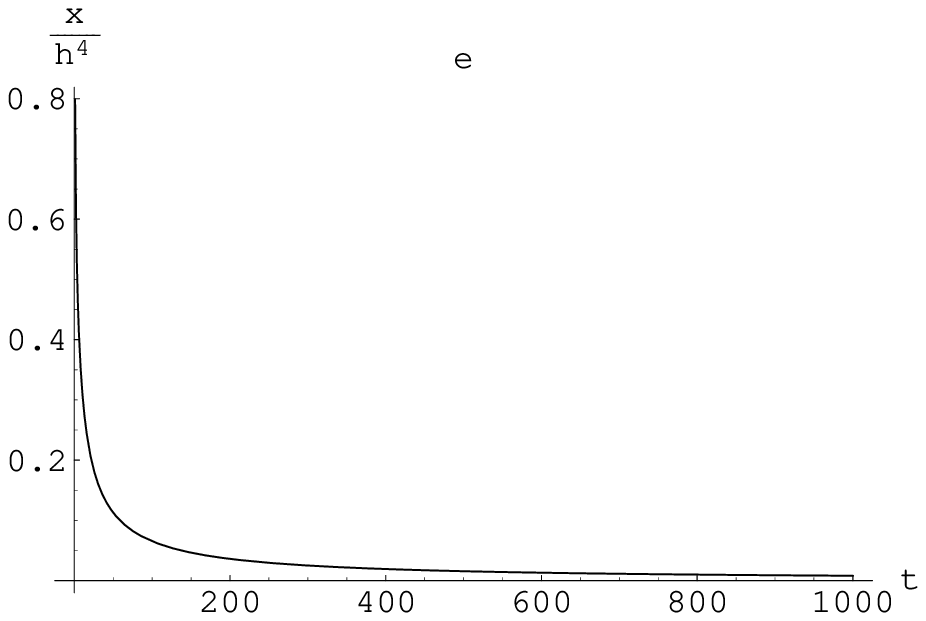} \epsfxsize6.5cm \epsfbox{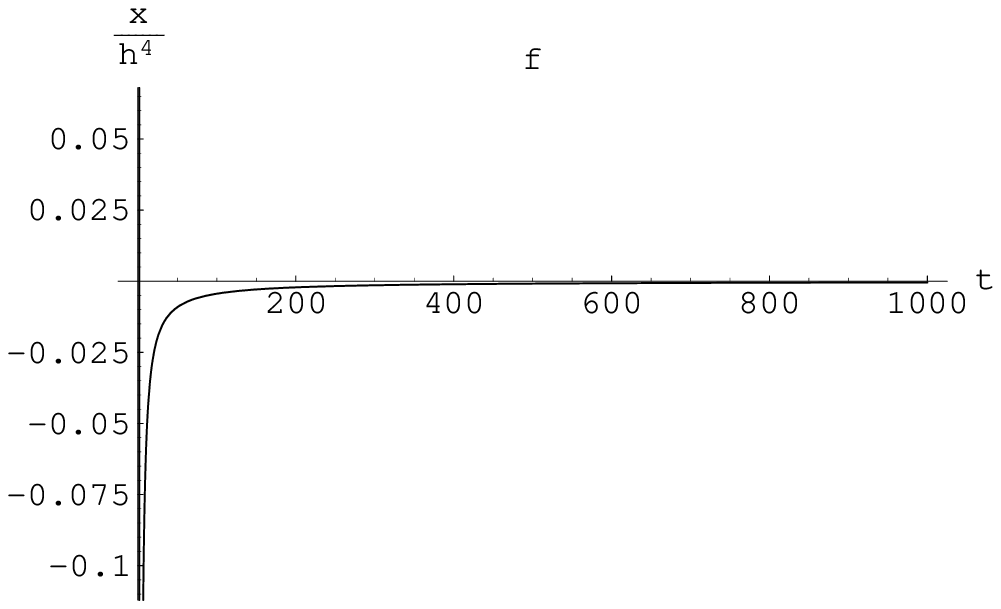}
\epsfxsize6.5cm  \epsfbox{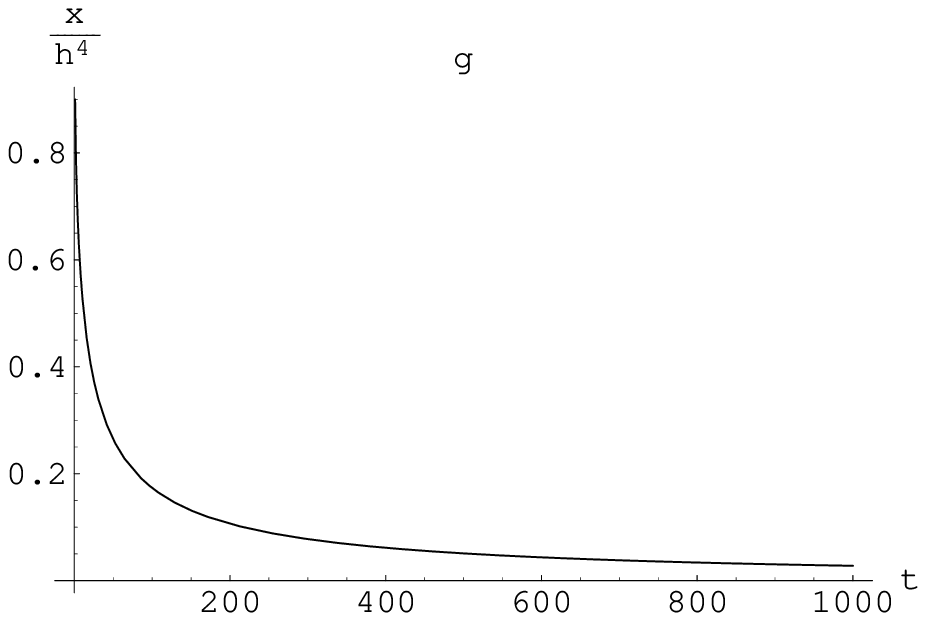} \caption{The dependence of
$x/h^4$ for different initial values $\phi_0$ and $\dot{\phi}_0$.} 
\label{Fig.1}
\end{figure}

In Fig. 2 we have plotted  all seven cases on the  time
range smaller than $1000$, 
displaying some additional details of the time evolution of
the value of $x/h^4$. It is clearly seen that, for $\dot \phi_0>0$
(cases b),d) and f)), transition from the kinetic-dominated
stage $x/h^4=w=1$ to the Chaplygin-like matter dominated stage
$x/h^4=w\approx 0$ goes through a short transient potential-dominated
regime $x/h^4=w=-1$.

\begin{figure}[h]
\epsfxsize6.5cm \epsfbox{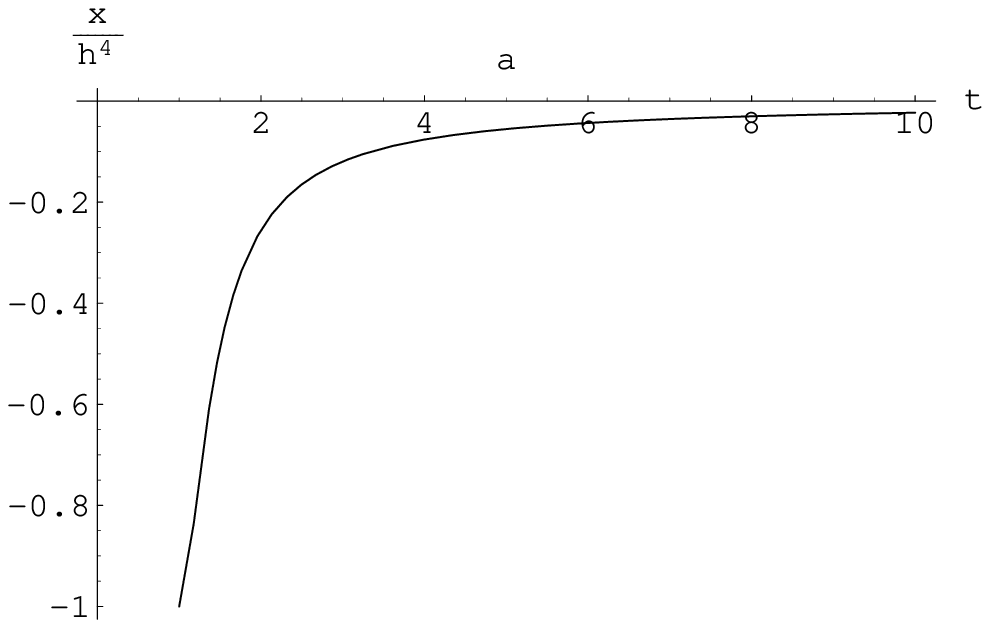} \epsfxsize6.5cm
\epsfbox{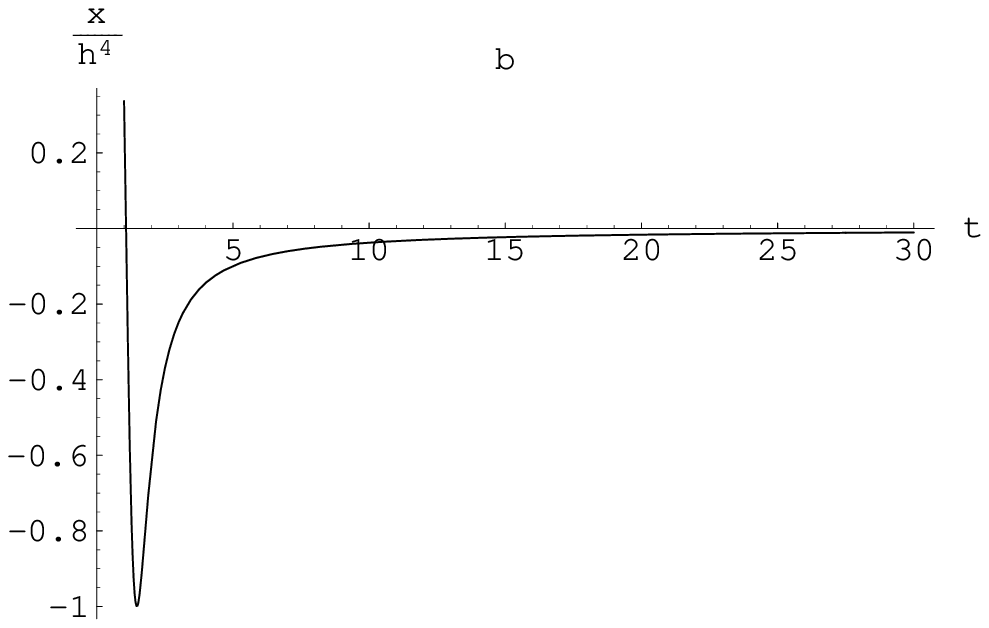} \epsfxsize6.5cm \epsfbox{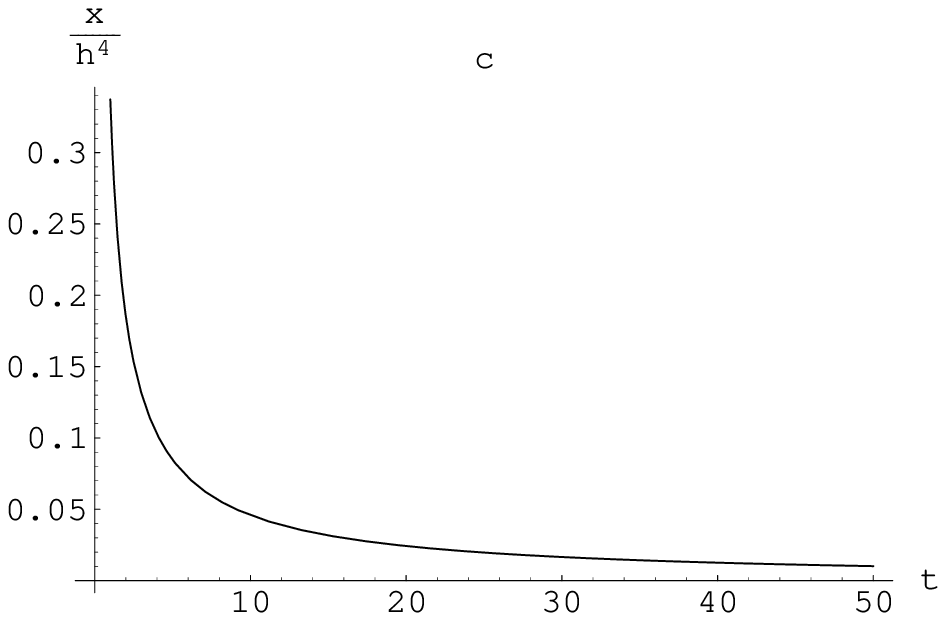}
\epsfxsize6.5cm \epsfbox{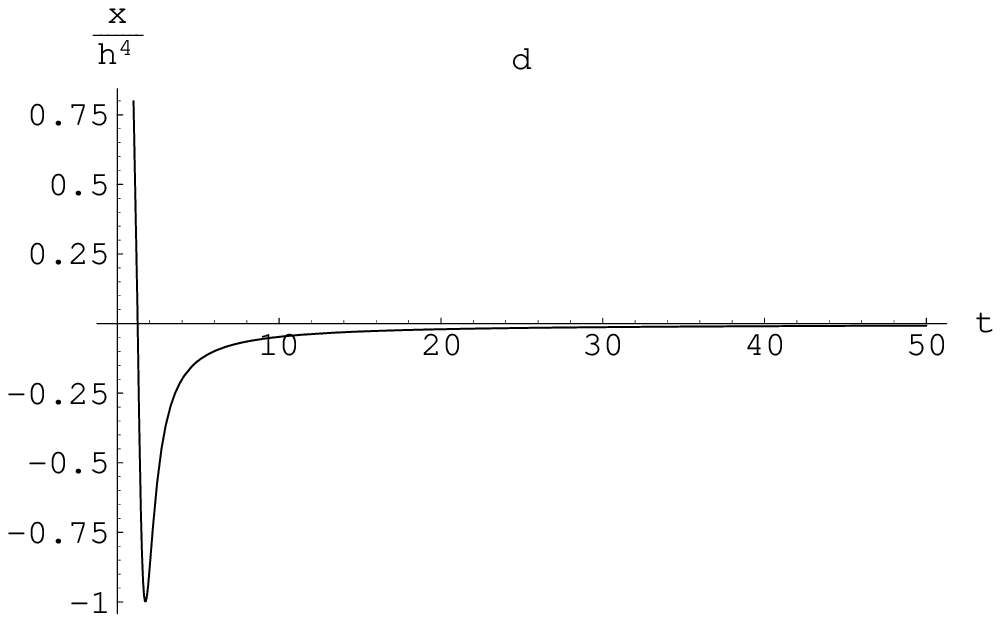} \epsfxsize6.5cm
\epsfbox{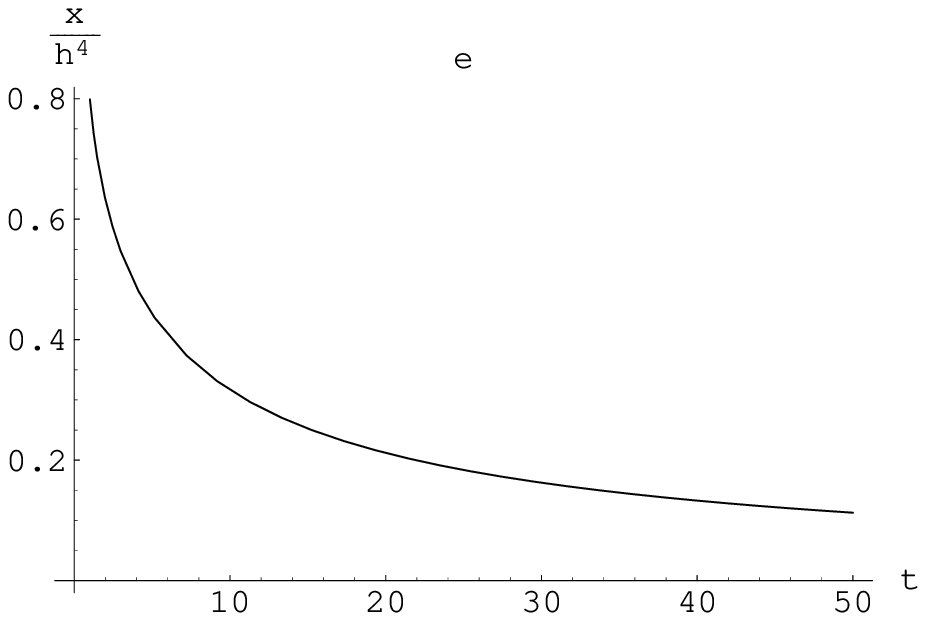} \epsfxsize6.5cm \epsfbox{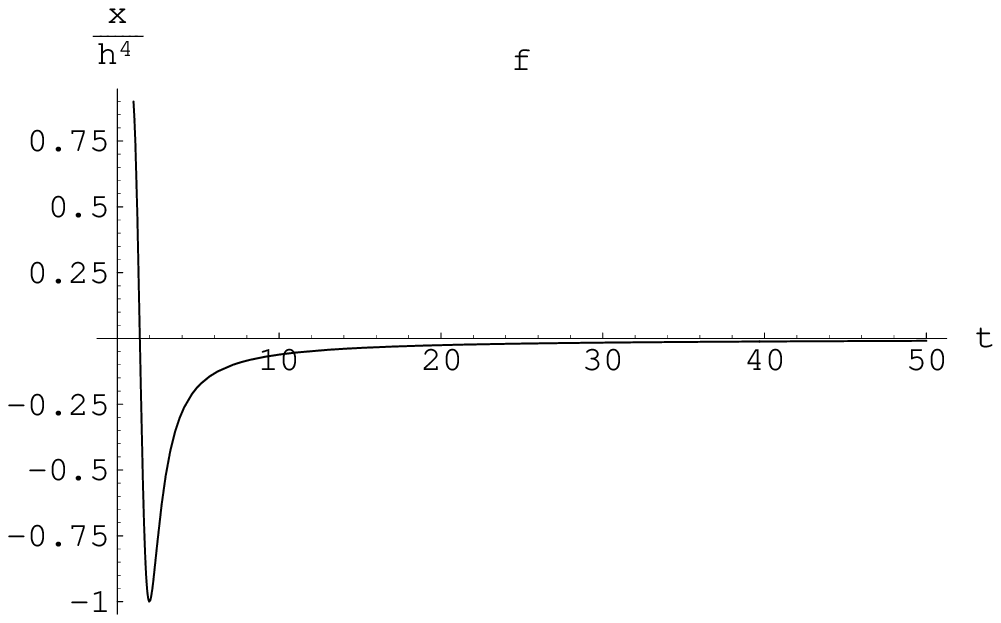}
\epsfxsize6.5cm  \epsfbox{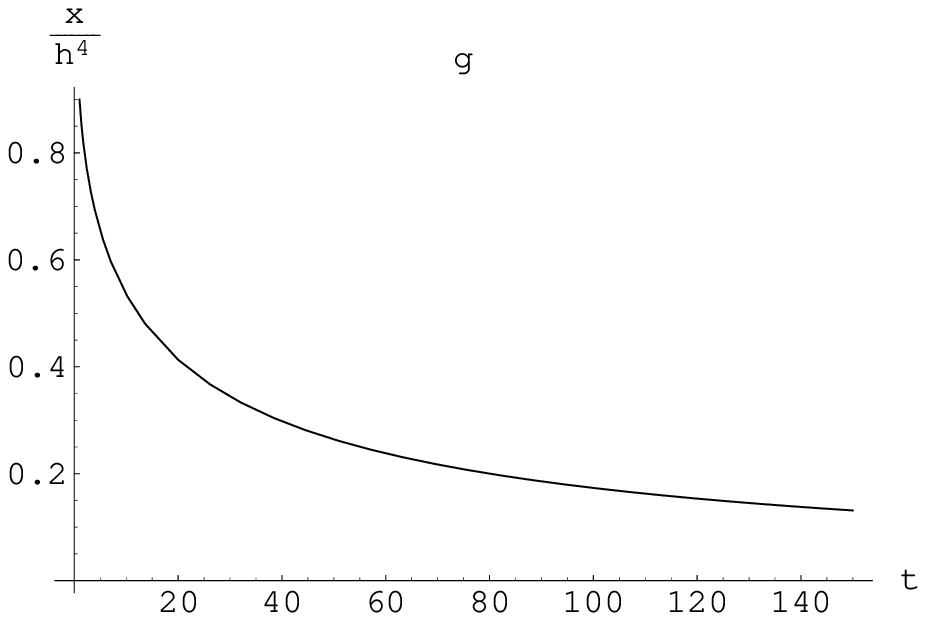} \caption{The dependence of
$x/h^4$ for different initial values $\phi_0$ and $\dot{\phi}_0$:
some details.} \label{Fig.2}
\end{figure}

We now return  to the calculation of a measure of initial
conditions near a cosmological singularity leading to a sufficiently
long matter dominated stage described by the scalar field model
(\ref{pot-Chap}) of the Chaplygin gas. To be concrete, now we
introduce two different probability measures in the space of initial
conditions. These measures are to be fixed on the basis of some
additional (usually
physical) assumptions. Depending on different assumptions, the
probability of a set of initial conditions may appear both large and
small for the same dynamical problem. For instance, in the case of a
massive scalar field in a closed FRW universe, the probability to
have a long quasi-de Sitter (inflationary) stage is very small if a
natural measure in the space of initial conditions is defined at the
moment of the maximal expansion of a non-singular solution with a
bounce \cite{St78}, while it is close to unity if initial conditions
are defined at the Planck energy density of the scalar field before
(or just at) the beginning of an expanding inflationary stage
\cite{BKGZ} (see also \cite{topor, pavl}).

We calculate the total probability to have condition
(\ref{cond}) satisfied for two different assumptions about a
measure in the space of initial conditions for $\phi$ and $\dot
\phi$ on the curve $\rho=1$ which corresponds to the Planck energy
scale. First, we assume the uniform measure in $\phi$, i.e.
\begin{equation}
d\mu = d\phi/2\phi_{max}. \label{uniform}
\end{equation}

For this purpose, we first calculate the total probability of the
opposite, undesired case when $|\phi_U|\le 8$. This clearly requires
$U\ll \dot\phi^2$ on the curve $\rho=1$. During the kinetic
dominated regime (\ref{free}), the undesired interval $|\phi_U|\le
8$ is shifted  either in the positive or in the negative direction
(depending on the sign of $\dot\phi$ in Eq. (\ref{free})) keeping
its width constant. The maximal value of $\phi$ on the $\rho=1$
curve is $\phi_{max}=\ln(1/\sqrt A)/3 \approx 93$.

Therefore, the total probability of undesired solutions is
\begin{equation}
P_- = {\ln z_{eq} \over \phi_{max}} \approx 9 \%~, \label{minus}
\end{equation}
while the total probability of the opposite, desired case with the
condition (\ref{cond}) being satisfied is
\begin{equation}
P_+ = 1 - P_- \approx 91 \%~. \label{plus}
\end{equation}

Now, following \cite{BKGZ}, we take the uniform measure on the
curve $\rho =1$ itself. Namely, we introduce the angular
parametrization
\begin{equation}
{\dot\phi^2\over 2}= \cos^2 \alpha~,~~~U(\phi)=\sin^2 \alpha~
\end{equation}
and assume the measure
\begin{equation}
d\mu = {d\alpha \over 2\pi}~.
\end{equation}
Hence, $U'd\phi=2\sin \alpha \cos \alpha d\alpha$. Note that for the
 massive scalar field considered in \cite{BKGZ}, the
second choice reduces to the first one for $|\alpha| \ll 1$
($|\phi| \ll \phi_{max}$). In our case, the total probability of
undesired solutions reads
\begin{equation}
P_-={1\over 2\pi}\int {d\alpha \over d\phi}d\phi = {3\over
2\pi}\int d\phi \sqrt U = {1\over \pi} \sqrt{U_{in,max}}~.
\end{equation}
Here, it is taken into account that only small regions around the
two points $\alpha =0,~\pi$ (with $U\ll \dot\phi^2$, but $|\phi|\gg
1$ already) contribute. $U_{in,max}$ is the maximal initial value
of the potential at the moment $t=t_{in}$ corresponding to $H^2 =
\rho = 1$ ($t_{in}=1/3$ for the kinetic dominated regime
(\ref{free})) for which the condition (\ref{cond}) is {\em not}
satisfied, so that $t_U\ge t_{eq}\sim H_0^{-1}z_{eq}^{-3/2}$.

The maximal initial value of $U$ on undesired trajectories is
saturated for $t_U=t_{eq}$ and with $\phi$ and $\dot{\phi}$ having
the opposite signs.
Then $U(\phi(t))\sim t_{eq}^{-2}$ at
$t=t_{eq}$, and it grows as $t$ increases $\sim t^{-\sqrt 2}$.
Therefore,
\begin{equation}
U_{in,max} \sim t_{eq}^{-2+\sqrt 2}t_{in}^{-\sqrt 2}\sim 10^{-32}
\end{equation}
and
\begin{equation}
P_-\sim 10^{-16}~, ~~~P_+ = 1 - {\cal O}(10^{-16})~.
\label{second}
\end{equation}
Thus, under the second assumption, undesired trajectories have
exceedingly small probability. The difference between (\ref{minus}),
(\ref{plus})  and (\ref{second}) illustrates the crucial dependence
of total probability values on assumptions made about a measure in
the space of initial conditions. However, in both cases, the
probability to reach the Chaplygin gas trajectory sufficiently early
to have the matter dominated stage in the Universe completely
described by the Chaplygin gas model is overwhelming, in sharp
contrast with the result of \cite{Perrotta}.

Thus, not only are the Chaplygin gas trajectories stable with
respect to those of the scalar field model with potential
(\ref{pot-Chap}), but also the probability to reach them
sufficiently early is close to unity for natural initial conditions
defined near a cosmological singularity.

In conclusion, we have shown that stability of trajectories of perfect
fluid cosmological models with respect to sets of trajectories of
more complicated field-theoretical models is a property which is not
always present, and sometimes it is present only for some pieces of
a trajectory under consideration. However, the Chaplygin gas
cosmological model provides an example of a model where entire
trajectories are stable.

AK and AS were also partially supported by the Russian Foundation
for Basic Research, grant No. 05-02-17450, by the Research
Programme "Astronomy" of the Russian Academy of Sciences and by
the scientific school grant No. 2338.2003.2 of the Russian Ministry
of Education and Science.

\end{document}